\documentclass[conference]{IEEEtran}
\IEEEoverridecommandlockouts
\usepackage{cite}
\usepackage{amsmath,amssymb,amsfonts}
\usepackage{graphicx}
\usepackage{textcomp}
\usepackage{xcolor}
\usepackage{url}
\usepackage{verbatim}
\usepackage{hyperref}
\usepackage{amsmath,amssymb,amsfonts}
\usepackage{algorithm}
\usepackage{algpseudocode}
\usepackage{bm}
\def\BibTeX{{\rm B\kern-.05em{\sc i\kern-.025em b}\kern-.08em
    T\kern-.1667em\lower.7ex\hbox{E}\kern-.125emX}}
\usepackage{tikz}
\usetikzlibrary{arrows.meta,positioning,shapes.geometric,calc}
\usepackage{booktabs}

\begin{document}

\title{Self-Verifying Anomaly Detection using Explainable AI for Cybersecurity of DER Networks}

\author{\IEEEauthorblockN{Damilola Popoola, Souradeep Bhattacharya, and Manimaran Govindarasu}
\IEEEauthorblockA{\textit{Department of Electrical and Computer Engineering} \\
\textit{Iowa State University, Ames, IA 50011}\\
Email: popson@iastate.edu, sbhatta@iastate.edu, gmani@iastate.edu}
}

\maketitle

\begin{abstract}
The rapid growth of Distributed Energy Resources (DERs) has significantly expanded the cyber-attack surface of modern power grids. Furthermore, increasing sophistication in attack techniques demands anomaly detection systems (ADS) that are accurate, interpretable, and reliable to support DER cybersecurity.  While ML-based ADS provide strong detection capabilities, their ``black-box" nature reduces operator trust and limits Security Operation Center's (SOC) ability to effectively interpret alerts and respond, highlighting the need for explainable Artificial Intelligence (XAI) to ensure transparency and operational confidence. This paper presents an XAI-based anomaly detection framework tailored for DER networks (ExCYDER). The proposed framework uses a self-verifying mechanism that validates ADS alerts to ensure trustworthy decision-making. ExCYDER combines LightGBM with SHAP to check whether each model decision aligns with its feature-attribution evidence, allowing the system to confirm that its internal reasoning is consistent and reliable. Experiments on a realistic DNP3 dataset achieved over 98\% detection accuracy, an average rule–SHAP consistency of 44.6\%, a SHAP latency of 14.5 ms per alert, and a confidence deviation within $\pm$5\%, demonstrating stable verification behavior with minimal computational overhead. The framework distinguished between coherent and inconsistent alerts without compromising detection accuracy, demonstrating that integrated verification within XAI-based ADS enhances interpretability, auditability, and operational robustness for DER-focused SOCs.

\end{abstract}


\section{Introduction}

Distributed Energy Resources (DERs) in modern power grids have transformed the energy landscape, enabling greater integration of renewable energy sources, enhanced grid flexibility, and improved energy efficiency \cite{javed_der_integration_2021}. However, this transformation has simultaneously introduced unprecedented cybersecurity challenges that threaten the stability, reliability, and security of critical energy infrastructure \cite{nrel_cybersecurity_2022}. As DER systems become increasingly interconnected, they present an expanded attack surface that malicious actors can exploit to compromise grid operations \cite{doe_cybersecurity_considerations_2022}. Traditional security approaches that depend on fixed, perimeter-based defenses are increasingly ineffective against modern, adaptive cyber threats. Although Artificial Intelligence (AI), Machine Learning (ML), and Deep Learning (DL)-based anomaly detection systems (ADS) provide strong real-time detection capabilities~\cite{sarker_ai_cybersecurity_2020}, their deployment in critical infrastructure is hindered by the inherent "black-box" behavior of many models~\cite{arrieta_explainable_ai_2020}. This opacity creates a trust deficit, as Security Operations Center (SOC) analysts require clear, justifiable explanations for security decisions.


Recognizing this gap, the field of Explainable Artificial Intelligence (XAI) has emerged to address the interpretability of AI models \cite{islam_explainable_ai_2022}, but mere explainability is not sufficient. True reliability and interpretability require not only an explanation but also verification of that explanation. Considering the critical nature of DER networks, this paper presents an XAI-based ADS framework for enhancing the cybersecurity of DER networks (ExCYDER). The primary contributions of this work are as follows: 
(1) development of ensemble learning-based ADS with an explainability mechanism for interpretable and reliable cybersecurity analytics tailored for DER SOC environments; 
(2) development of an explainability-driven verification process that validates model decisions through rule–feature consistency analysis, enhancing the transparency, confidence, and auditability of SOC-level anomaly investigation; and 
(3) real-time testbed-based evaluation of ExCYDER using realistic DER datasets to demonstrate its detection accuracy, verification stability, and interpretability performance.

The remainder of this paper is organized as follows. 
Section~II provides background on DER communication and cybersecurity challenges. 
Section~III reviews related work in anomaly detection and explainable AI. 
Section IV presents the proposed ExCYDER methodology, including the edge-cloud workflow, explainability mechanisms, and verification logic. 
Section~V describes the experimental setup, performance metrics, and evaluation results. 
Finally, Section~VI concludes the paper and outlines directions for future work.

\begin{figure*}[t]
\centerline{\includegraphics[width=\textwidth, height = 9cm]{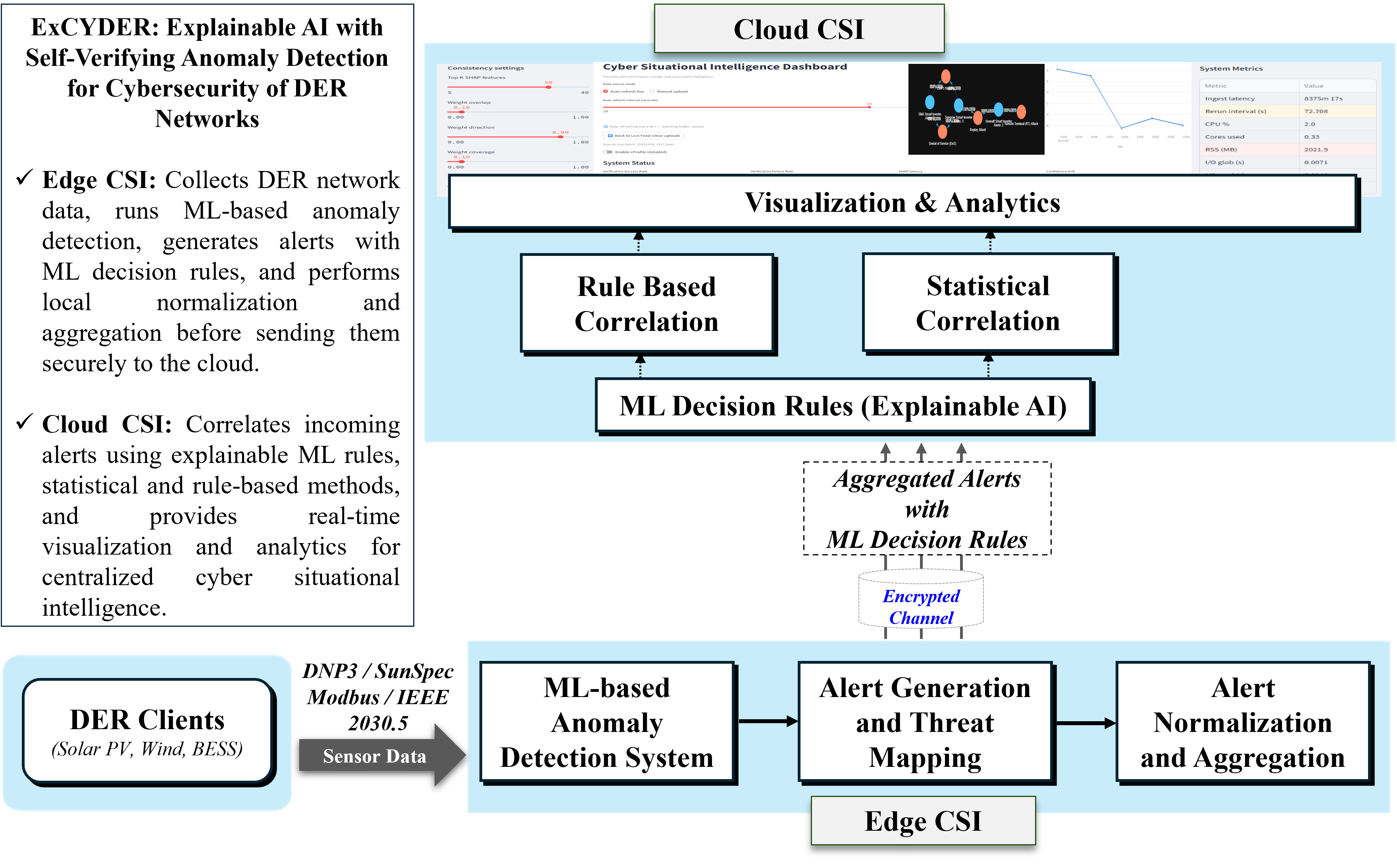}}
\vspace{-0.2cm}
\caption{ExCYDER workflow from edge-based anomaly detection and alert aggregation to cloud-level verification, correlation, and visualization}
\label{fig:framework}
\vspace{-0.5cm}
\end{figure*}

\section{Background on DER Cybersecurity}

\subsection{DER Communication Architecture}

The DER communication structure in this work adopts the three-tier DER architecture described in our earlier framework, CySIDER~\cite{popoola_cysider_2025}. It comprises: (1) \textit{DER clients and devices} such as solar PVs, battery systems, and EVs that regulate local generation and storage; (2) an \textit{edge intelligence layer} that hosts anomaly detection and secure telemetry exchange through Edge Intelligent Devices (EIDs); and (3) a \textit{utility and aggregator layer} that coordinates distributed assets and manages grid interaction through the DER Management System (DERMS). Standardized communication protocols (e.g., DNP3, Modbus, IEC~61850, IEEE~2030.5) enable interoperability across heterogeneous devices but are limited by weak or absent authentication and encryption mechanisms.

\subsection{Cybersecurity Challenges}

As detailed in~\cite{popoola_cysider_2025}, the growing digital interconnection of DER assets has significantly expanded the grid’s attack surface. Protocols such as DNP3 and SunSpec Modbus were not originally designed with built-in cryptographic safeguards, exposing field devices, including smart inverters, EIDs, and vendor gateways to threats such as replay, spoofing, denial-of-service, and data manipulation. These vulnerabilities, combined with fragmented alert streams and opaque ML models, hinder timely detection and verification within SOCs. Consequently, modern DER cybersecurity requires explainable, verifiable, and SOC-integrated anomaly detection frameworks such as ExCYDER to ensure reasoning transparency, incident traceability, and operational resilience in distributed power networks.


\subsection{Need for Explainability in ML based Anomaly Detection}

DER networks increasingly require transparent anomaly detection models rather than traditional “black box” approaches. Although DL and ensemble methods achieve high accuracy, their opacity limits operator trust and slows incident response. XAI improves confidence by revealing the features behind each prediction and supporting debugging and auditability~\cite{arrieta_explainable_ai_2020,islam_explainable_ai_2022}. However, post hoc explanations can be inconsistent or misaligned with model behavior, reducing their reliability in high stakes settings. This motivates the need for approaches that not only explain model outputs but also verify the consistency and reliability of those explanations.

\subsection{Related Work}

Traditional anomaly detection in power systems has relied on statistical models and threshold-based alarms, which often fail to cope with the complexity and high-dimensional data streams of modern DER networks. Recent research has shifted toward ML and DL techniques, which offer superior capability in identifying subtle attack patterns. For instance, ensemble methods and neural networks have been explored for detecting data injection and denial-of-service attacks in smart grid environments~\cite{khraisat_survey_intrusion_2019,villegas_integrating_2025}. While these models achieve high detection accuracy, their ``black-box'' nature poses a critical challenge for SOC analysts, who require clear, justifiable evidence to triage and respond to alerts in real time.

The field of XAI has emerged to bridge the gap between high model performance and interpretability. In the context of cybersecurity, XAI enhances transparency, fosters trust, and facilitates collaboration between human analysts and AI tools~\cite{zhang_explainable_2022,rjoub_survey_2023}. 
XAI methods such as LIME and SHAP have been applied to energy systems to improve model transparency \cite{paul_explainable_ai_energy_2024,al_ashmali_ensemble_learning_2023}. These techniques assist analysts in understanding model predictions and debugging anomalous behavior. Recent studies also highlight the importance of interpreting power grid anomalies from a human centered perspective \cite{wang_interpretability_2025}.

However, these XAI applications provide only post hoc explanations and do not verify the internal consistency of model reasoning. This limits their usefulness in SOC environments where rapid and reliable interpretation is required.



\section{Proposed Methodology}

To address the limitations identified in existing explainable anomaly detection approaches, this work proposes the ExCYDER framework, designed to support DER-specific SOC operations as shown in Fig~\ref{fig:framework}. ExCYDER integrates ensemble-based anomaly detection with a self-verifying explainability mechanism that internally validates each model decision through consistency analysis. The framework operates across two hierarchical layers: (1) an \emph{edge intelligence layer} that performs real-time anomaly detection and alert generation near DER devices, and (2) a \emph{cloud intelligence layer} that aggregates, verifies, and visualizes alerts to support analyst interpretation and situational intelligence.

\subsection{ML Based Anomaly Detection}

At the edge intelligence layer, ExCYDER adopts the ML-Based ADS framework developed in~\cite{bhattacharya_mlads_2024}. The adopted ML ADS pipeline performs systematic data preprocessing, including label encoding of categorical variables, feature selection to remove redundant or low variance attributes, and normalization using Min–Max scaling, followed by dataset balancing through the Synthetic Minority Oversampling Technique (SMOTE). The balanced dataset is then divided into 70\% training and 30\% testing subsets for model evaluation.



\subsection{Model Explainability using SHAP and Decision Rules}

ExCYDER derives its interpretability from the combined use of model decision rules and SHapley Additive exPlanations (SHAP). The decision rules extracted from the LightGBM model capture the logical path leading to each prediction, while SHAP quantifies the contribution of each feature to that decision. Together, they represent the model’s decision logic and supporting evidence. The alignment between these two components forms the foundation of ExCYDER’s verification process, enabling the system to evaluate how consistently the model’s explanations reflect its internal decision logic.

\subsection{Cloud-Level Verification}

At the cloud layer, ExCYDER verifies each aggregated alert by cross-checking the model’s rule-based decision path with corresponding SHAP feature attributions. This determines whether the model’s explanation (evidence) aligns with its internal decision logic (claim), producing a consistency score for each alert before it is presented on the SOC dashboard.

The verification process employs a multi-term consistency metric inspired by recent studies on explanation fidelity and faithfulness~\cite{bhatt2020evaluating}. The metric integrates three components: \textit{Overlap} $(O)$, \textit{Direction} $(D)$, and \textit{Coverage} $(C)$, which jointly measure the agreement between decision rules and SHAP attributions. Overlap captures shared features between the decision path and top-$K$ SHAP features, Direction checks whether SHAP contributions follow the same causal trend as rule inequalities, and Coverage quantifies the SHAP importance explained by shared features. These terms form a unified consistency score optimized for SOC-level interpretability.

\begin{figure}[t]
\centering
\begin{tikzpicture}[
  node distance=3mm and 2mm,
  >=Latex,
  font=\scriptsize,
  box/.style={draw, rounded corners=1pt, align=center, inner sep=3pt, fill=gray!7},
  dia/.style={draw, diamond, aspect=2.1, align=center, inner sep=2pt, fill=gray!7}
]
\node[box] (in) {Input $x$; model $M$};
\node[box, below=of in] (rule) {Rule extraction \\ $r\!=\!R(M,x)$, $\mathcal F_{\text{rule}}$};
\node[box, below=of rule] (shap) {SHAP attribution \\ $S(x)$, top-$K$ $\mathcal F_{\text{shap}}$};
\node[box, below=of shap] (cons) {Compute consistency: $O, D, C$ \\ $\text{Cons}=w_{ov}O+w_{dir}D+w_{cov}C$};
\node[box, below=of cons] (blend) {Blend confidence \\ $\hat p=\alpha p+(1-\alpha)\,100\cdot \text{Cons}$};
\node[dia, below=of blend] (th) {$\text{Consistency}\ge \theta$?};
\node[box, left=10mm of th] (unv) {Unverified alert \\Output $y,\hat p$};
\node[box, right=10mm of th] (ver) {Verified alert \\Output $y,\hat p$};

\draw[->] (in) -- (rule);
\draw[->] (rule) -- (shap);
\draw[->] (shap) -- (cons);
\draw[->] (cons) -- (blend);
\draw[->] (blend) -- (th);
\draw[->] (th) -- node[above,sloped]{Yes} (ver);
\draw[->] (th) -- node[above,sloped]{No} (unv);
\end{tikzpicture}
\vspace{-0.2cm}
\caption{ExCYDER workflow for explainability-driven alert verification}

\label{fig:excyder_flow}
\vspace{-0.3cm}
\end{figure}
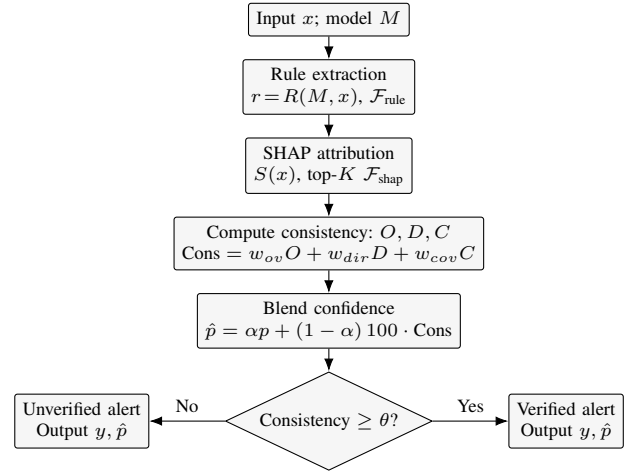

As illustrated in Fig.~\ref{fig:excyder_flow}, the system performs rule extraction, SHAP attribution, consistency computation, and confidence blending. Alerts meeting the verification threshold $\theta$ are labeled \emph{verified}, while others are flagged as \emph{unverified} for additional analyst review.


\begin{figure}[t]
  \centering
  \includegraphics[width=\linewidth, height = 8cm]{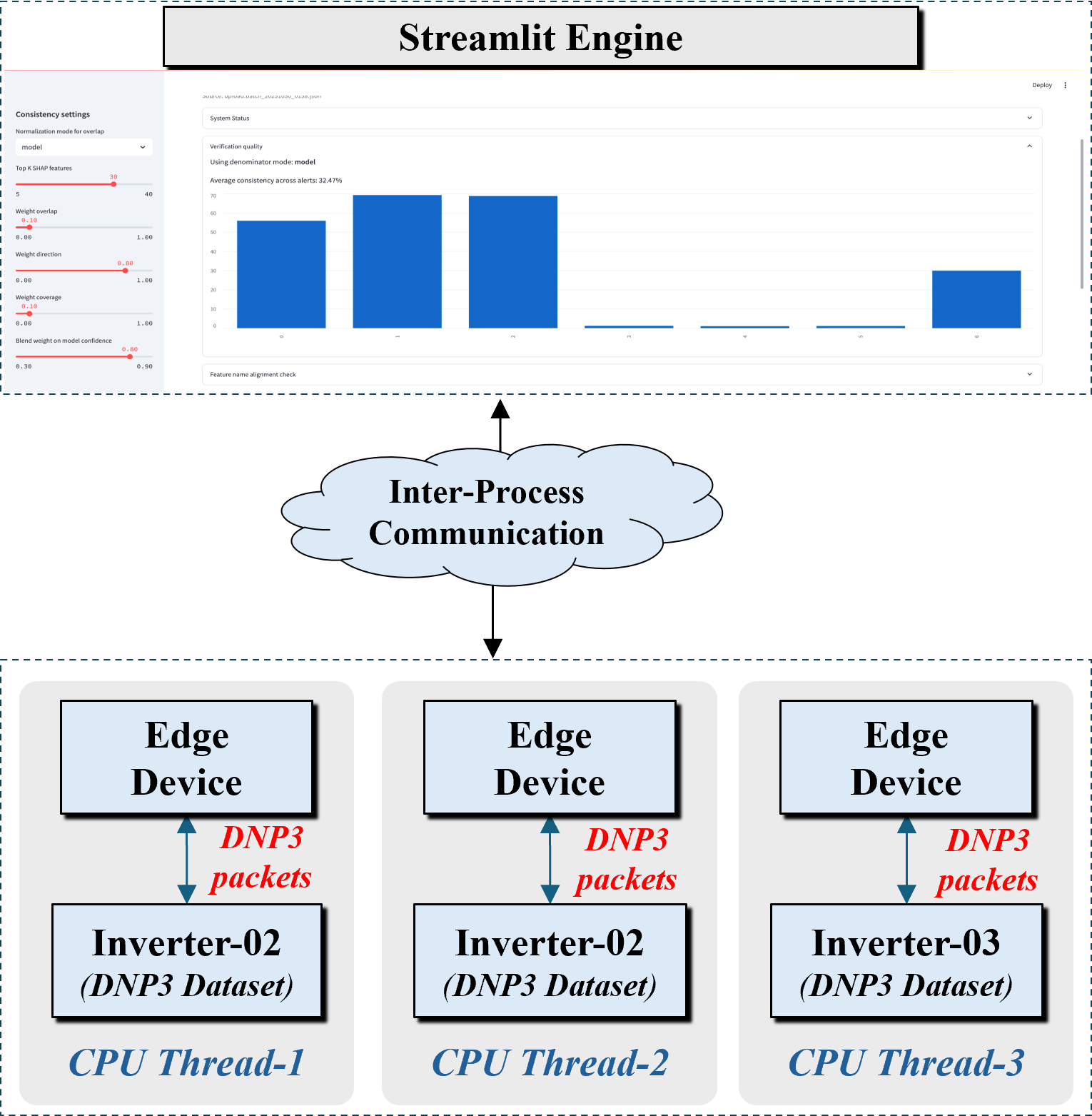}
  \vspace{-0.5cm}
  \caption{Experimental setup showing communication between the DER devices, edge devices and the cloud analytics layer}
  \label{fig:exp_framework}
  \vspace{-0.5cm}
\end{figure}

\section{Experimental Evaluation}

\subsection{Experimental Setup}

Fig.~\ref{fig:exp_framework} illustrates the experimental setup, which emulates a distributed SOC-oriented cybersecurity environment for DER networks. Three virtual DER devices were implemented in Python~3.13 on an Apple M4 system (16~GB RAM, macOS~15.5), each replaying labeled DNP3 traffic in real time to ExCYDER’s edge anomaly detection module. The edge layer performs local inference and aggregates alerts into JSON batches transmitted to the cloud every 60 seconds \cite{11133877}. The cloud processes each batch on the same interval for verification and visualization, providing near-real-time situational awareness without excessive update frequency. Verified alerts are displayed on the SOC dashboard with their classifications, adjusted confidence values, and consistency scores to support analyst decision-making.

\begin{table}[t]
\caption{Performance Evaluation Metrics}
\vspace{-0.2cm}
\centering
\renewcommand{\arraystretch}{1.05}
\setlength{\tabcolsep}{4pt}
\begin{tabular}{p{2.8cm}p{5.4cm}}
\toprule
\textbf{Metric} & \textbf{Description / Formula} \\
\midrule
\multicolumn{2}{l}{\textbf{(a) Verification and Explainability Metrics}} \\
\midrule
Verification Success rate (\%) & Ratio of alerts whose consistency exceeds threshold $\theta$; reflects reasoning alignment. \\
Verification Failure Rate (\%) & Fraction of alerts that did not pass the verification threshold at $\theta = 50\%$; represents unverified or low-consistency alerts. \\
Average consistency (\%) & Mean rule–SHAP consistency across all alerts; measures interpretability and model coherence. \\
Confidence deviation (\%) & Average difference between baseline and verified confidence; quantifies verification stability. \\
Average SHAP time (ms) & Mean computation time for SHAP feature attribution; represents explainability overhead. \\
\midrule
\multicolumn{2}{l}{\textbf{(b) System Performance Metrics}} \\
\midrule
Ingest latency (s) & Delay between alert batch creation and visualization; end-to-end data transfer efficiency. \\
Rerun interval (s) & Time between dashboard refresh cycles; ensures alignment with the 60 s verification schedule. \\
CPU utilization (\%) & Processor load per refresh; values above 70\% indicate heavy computation overhead. \\
RSS memory (MB) & Resident memory footprint of the Streamlit process; tracks memory usage. \\
I/O glob (s) & Time spent listing all JSON batch files in the directory; measures disk or network latency. \\
I/O read (s) & Time to open and read the latest JSON file. \\
JSON parse (s) & Time to deserialize JSON content into Python objects; reflects data volume and parsing cost. \\
\bottomrule
\end{tabular}
\vspace{-0.5cm}
\label{tab:excyder_metrics_final}
\end{table}

\begin{figure}[t]
\centering
\includegraphics[width=0.50\textwidth, height = 9cm]{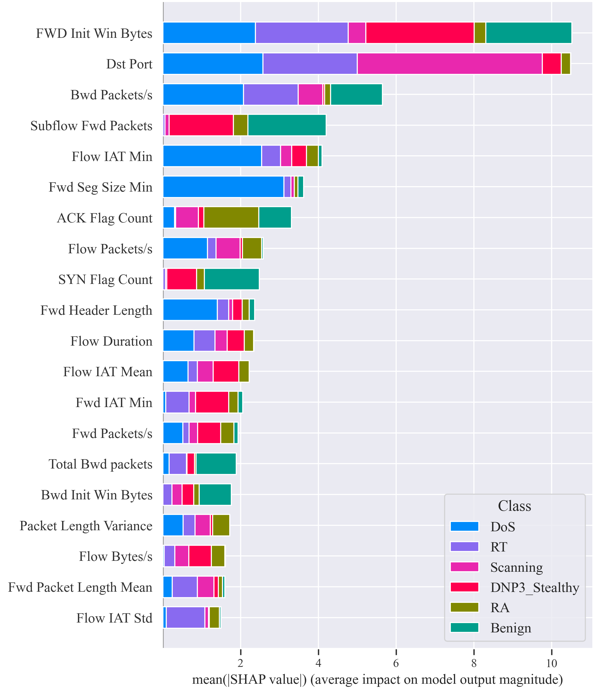}
\vspace{-0.3cm}
\includegraphics[height = 9cm]{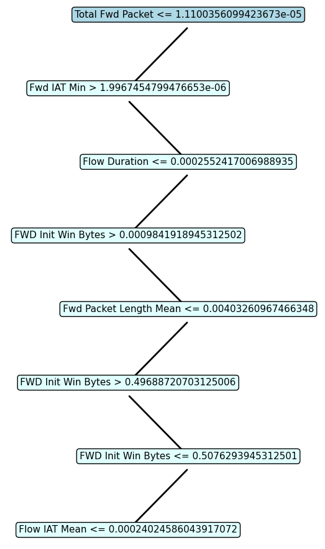}
\caption{(Top) Global SHAP summary plot for the LightGBM showing feature importance across alert classes. (Bottom) Decision rule path for DoS illustrating shared classification logic}
\label{fig:shap_bar_lightgbm}
\vspace{-0.5cm}
\end{figure}

\subsection{DER Dataset}

This work uses the Iowa State University DER DNP3 dataset \cite{abdelkhalek_der_ads}, which provides labeled DNP3 communication records collected from a DER network environment. The dataset defines six traffic classes that capture both normal operation and representative cyberattacks within DER communication: Benign Traffic, Denial of Service, Remote Terminal Attack, Scanning Attack, DNP3 Stealthy Attack, and Replay Attack. These labels form a reliable basis for supervised training and evaluation of the anomaly detection system by distinguishing routine behavior from malicious activity.


\subsection{Results and Discussion}

The evaluation focused on validating ExCYDER’s anomaly detection performance, explanation fidelity, and verification consistency under a SOC oriented simulation environment. A total of 250 DER communication samples were simulated over a 7 minute interval, generating 121 malicious events that included DNP3 Stealthy, Denial of Service (DoS), and Replay attack scenarios, three prominent inverter layer vulnerabilities documented in~\cite{forescout_sundown_2025}. These events were aggregated into seven JSON batches using the 60 second edge batching interval. Each batch was processed by ExCYDER’s cloud level verification engine, where rule based decision paths and SHAP feature attributions were cross validated before being visualized on the SOC dashboard for analyst interpretation.

\textbf{Machine Learning Detection Performance:} 
As established in our earlier work \cite{popoola_cysider_2025}, the ML based ADS demonstrates high detection reliability on DER communication traffic. That evaluation assessed LightGBM, XGBoost, and Random Forest using standard classification metrics, including Precision (correct identification of attacks among all predicted attacks), Recall (correct detection of attacks among all actual attacks), F1 score (the harmonic balance of Precision and Recall), and the False Negative Rate (missed attacks). LightGBM achieved the strongest performance with more than 99 percent Precision, Recall, and F1 score, and the lowest False Negative Rate of 0.51 percent, indicating minimal missed attacks and clear separation between benign and malicious traffic. XGBoost and Random Forest followed with higher False Negative Rate values of 0.88 percent and 5.10 percent, respectively. Based on these validated results, LightGBM is adopted as the operational anomaly detector in ExCYDER.

\textbf{Explainability Analysis:} Fig.~\ref{fig:shap_bar_lightgbm} presents the global SHAP summary of feature contributions across all alert classes and decision-path for DOS attack. Attributes such as \textit{FWD Init Win Bytes}, \textit{Dst Port}, and \textit{Bwd Packets/s} were the most influential features shaping model decisions, reflecting their high correlation with protocol-level activity. The decision-path illustrates how specific thresholds in these features lead to the classification of a DoS event. These insights confirm that the model captures meaningful traffic behaviors rather than relying on spurious correlations, thereby improving interpretability.

\textbf{Verification Outcomes:} Consistency scores varied across alerts due to differences in the model’s decision paths and feature attributions, but the overall average remained $44.6\%$, indicating moderate alignment between rule driven reasoning and SHAP attributions. Overall, the verification module achieved a $54.5\%$ success rate, with alerts below the $50\%$ threshold filtered as unverified ($45.5\%$). The confidence deviation stayed within $\pm 5\%$, confirming that verification preserved classifier reliability. The average SHAP computation time was $14.5$~ms per alert, demonstrating the feasibility of explainability in near real time SOC contexts.


\textbf{Operational Stability:} During continuous runtime, system metrics confirmed the dashboard’s responsiveness and computational efficiency. The average ingest latency remained below 10~s, and the rerun interval synchronized with the 60-second aggregation cycle. CPU utilization remained under 70\%, and resident memory (RSS) stayed below 900~MB, confirming the lightweight nature of ExCYDER’s verification loop. These metrics validate the framework’s suitability for continuous SOC operation without performance degradation.

\section{Conclusion}

The results demonstrate that ExCYDER effectively integrates high-accuracy ML detection with interpretable and verifiable reasoning for DER-focused SOC environments. By combining rule-based reasoning, SHAP feature attribution, and consistency-driven verification, the framework transforms post-hoc explanations into actionable and audit-ready intelligence. This internal validation mechanism also reduces redundant or low-confidence alerts, supports transparent triage, and enhances analyst trust in automated decision support. Future work include: (1) real-time deployment of ExCYDER within operational DER networks to evaluate performance, latency, and scalability under production conditions and, (2) development of interactive interfaces that integrate operator expertise with AI-driven insights, improving human–AI collaboration for SOC decision-making and threat intelligence.

\section*{Acknowledgement}
This work is funded in part by the US DOE Cybersecurity, Energy Security, and Emergency Response (CESER) Award Number DE-CR0000049.

\bibliographystyle{IEEEtran}
\bibliography{references}

@inproceedings{javed_der_integration_2021,
  author    = {A. H. Javed and P. H. Nguyen and J. Morren and J. G. H. Slootweg},
  title     = {Review of operational challenges and solutions for {DER} integration with distribution networks},
  booktitle = {Proc. 56th Int. Universities Power Eng. Conf. (UPEC)},
  year      = {2021},
  pages     = {1--6},
  address   = {Middlesbrough, UK},
  month     = aug,
  doi       = {10.1109/UPEC50034.2021.9548198}
}

@techreport{nrel_cybersecurity_2022,
  author      = {{National Renewable Energy Laboratory}},
  title       = {The distributed energy resource cybersecurity framework},
  institution = {National Renewable Energy Laboratory},
  number      = {NREL/TP-5D00-82577},
  address     = {Golden, CO, USA},
  year        = {2022},
  url         = {https://www.nrel.gov/docs/fy22osti/82577.pdf}
}

@techreport{doe_cybersecurity_considerations_2022,
  author      = {{U.S. Department of Energy}},
  title       = {Cybersecurity Considerations for Distributed Energy Resources on the {U.S.} Electric Grid},
  institution = {U.S. Department of Energy},
  address     = {Washington, DC, USA},
  year        = {2022},
  month       = oct
}

@article{sarker_ai_cybersecurity_2020,
  author  = {I. H. Sarker and A. S. M. Kayes and S. Badsha and H. Alqahtani and P. Watters and A. Ng},
  title   = {Cybersecurity data science: An overview from machine learning perspective},
  journal = {J. Big Data},
  volume  = {7},
  number  = {1},
  pages   = {1--29},
  year    = {2020},
  month   = dec,
  doi     = {10.1186/s40537-020-00318-5}
}

@article{arrieta_explainable_ai_2020,
  author  = {A. B. Arrieta and others},
  title   = {Explainable artificial intelligence ({XAI}): Concepts, taxonomies, opportunities and challenges toward responsible {AI}},
  journal = {Inf. Fusion},
  volume  = {58},
  pages   = {82--115},
  year    = {2020},
  month   = jun,
  doi     = {10.1016/j.inffus.2019.12.012}
}

@article{islam_explainable_ai_2022,
  author  = {S. R. Islam and W. Eberle and S. K. Ghafoor and M. Ahmed},
  title   = {Explainable artificial intelligence approaches: A survey},
  journal = {IEEE Access},
  volume  = {9},
  pages   = {101974--102001},
  year    = {2021},
  doi     = {10.1109/ACCESS.2021.3097166}
}

@inproceedings{paul_explainable_ai_energy_2024,
  author    = {S. Paul and others},
  title     = {Demystifying cyberattacks: Potential for securing energy systems with explainable {AI}},
  booktitle = {Proc. 2024 Workshop Comput., Netw. Commun. (CNC)},
  year      = {2024},
  pages     = {430--435},
  address   = {Las Vegas, NV, USA}
}

@article{al_ashmali_ensemble_learning_2023,
  author  = {I. G. Al-Ashmali and others},
  title   = {Enhancing network security: Robust anomaly detection with ensemble learning and explainable {AI}},
  journal = {Comput. Secur.},
  volume  = {125},
  pages   = {103118},
  year    = {2023},
  month   = feb,
  doi     = {10.1016/j.cose.2022.103118}
}

@article{khraisat_survey_intrusion_2019,
  author  = {A. Khraisat and I. Gondal and P. Vamplew and J. Kamruzzaman},
  title   = {Survey of intrusion detection systems: Techniques, datasets and challenges},
  journal = {Cybersecurity},
  volume  = {2},
  number  = {1},
  pages   = {1--22},
  year    = {2019},
  month   = dec,
  doi     = {10.1186/s42400-019-0038-7}
}

@article{villegas_integrating_2025,
  author  = {W. Villegas-Ch and A. Jaramillo-Alcázar and A. M. Navarro and A. Mera-Navarrete},
  title   = {Integrating Explainable Artificial Intelligence in Anomaly Detection for Threat Management in E-Commerce Platforms},
  journal = {IEEE Access},
  volume  = {13},
  pages   = {29830--29846},
  year    = {2025}
}

@article{zhang_explainable_2022,
  author  = {Z. Zhang and H. A. Hamadi and E. Damiani and C. Y. Yeun and F. Taher},
  title   = {Explainable Artificial Intelligence Applications in Cyber Security: State-of-the-Art in Research},
  journal = {IEEE Access},
  volume  = {10},
  pages   = {93310--93333},
  year    = {2022}
}

@article{rjoub_survey_2023,
  author  = {G. Rjoub and J. Bentahar and O. Abdel Wahab and R. Mizouni and A. Mourad},
  title   = {A Survey on Explainable Artificial Intelligence for Cybersecurity},
  journal = {IEEE Transactions on Network and Service Management},
  volume  = {20},
  issue   = {3},
  pages   = {3241--3260},
  year    = {2023},
  doi     = {10.1109/TNSM.2023.3282740}
}

@inproceedings{wang_interpretability_2025,
  author  = {D. Wang and T. C. Chen and K. Mahapatra},
  title   = {Interpretability, Explainability and Trust Metrics in Anomaly Detection Method for Power Grid Sensors},
  booktitle = {2025 IEEE PES Grid Edge Technologies Conference and Exposition (Grid Edge)},
  year    = {2025},
  address = {San Diego, CA, USA},
  month   = {April}
}

@inproceedings{bhatt2020evaluating,
  author    = {Bhatt, Umang and Weller, Adrian and Moura, Jose},
  title     = {Evaluating and Aggregating Feature-based Model Explanations},
  booktitle = {Proceedings of the 2020 Conference on Empirical Methods in Natural Language Processing (EMNLP)},
  pages     = {4550--4561},
  year      = {2020}
}

@techreport{forescout_sundown_2025,
  author       = {Forescout Research Labs},
  title        = {SUNDOWN: Exploiting Design Flaws in Industrial and Solar Power Communication Protocols},
  institution  = {Forescout Technologies, Inc.},
  year         = {2025},
  month        = {March},
  pages        = {1--44},
  note         = {Technical Vulnerability Research Report},
  url          = {https://www.forescout.com/resources/sun-down-research-report/},
  address      = {San Jose, CA, USA},
  doi          = {10.5281/zenodo.14875203}
}

@inproceedings{popoola_cysider_2025,
  author    = {Damilola Popoola and Souradeep Bhattacharya and Manimaran Govindarasu},
  title     = {CySIDER: Cybersecurity Situational Intelligence Framework for {DER} Networks},
  booktitle = {2025 Resilience Week (RWS)},
  address   = {National Harbor, MD, USA},
  year      = {2025},
  note      = {(Accepted)}
}

@inproceedings{bhattacharya_mlads_2024,
  author    = {Souradeep Bhattacharya and Nazmus Saqib and Manimaran Govindarasu},
  title     = {ML-based Anomaly Detection System for {IEC} 61850 Communication in Substations},
  booktitle = {2024 IEEE Power \& Energy Society General Meeting (PESGM)},
  address   = {Seattle, WA, USA},
  year      = {2024},
  doi       = {10.1109/PESGM51994.2024.10688773}
}

@INPROCEEDINGS{11133877,
  author={Bhattacharya, Souradeep and Govindarasu, Manimaran},
  booktitle={2025 34th International Conference on Computer Communications and Networks (ICCCN)}, 
  title={Real-time Cybersecurity Situational Awareness Framework for Agriculture Machinery-based IoT Networks}, 
  year={2025},
  volume={},
  number={},
  pages={1-9},
  doi={10.1109/ICCCN65249.2025.11133877}}

@misc{abdelkhalek_der_ads,
  author       = {Moataz AbdElKhalek},
  title        = {{DER\_ML\_ADS: Machine Learning-Based Anomaly Detection for DER Networks}},
  year         = {2022},
  note         = {Accessed: 2025-07-09},
  url          = {https://github.com/Moataz-AbdElKhalek/DER\_ML\_ADS/tree/main/dataset/DNP3}
}

\end{document}